\documentclass[twoside]{article}
\usepackage{a4wide}
\usepackage{amssymb, amsthm, amsmath}
\usepackage{pstricks}
\def\pstlw{.8pt}  

\newcommand{\fra}[2]{\textstyle{\frac{#1}{#2}}}
\newcommand{\beqn}{\begin{eqnarray}\begin{aligned}}
\newcommand{\eqn}{\end{aligned}\end{eqnarray}}

\begin{document}
\begin{titlepage}
\vskip 1.6in
\begin{center}
{\Large {\bf Using the tangle: a consistent construction of phylogenetic distance matrices for quartets}}
\end{center}

\normalsize
\vskip .4in

\begin{center}
J G Sumner$^*$ and P D Jarvis$^{\dagger}$  
\par \vskip .1in \noindent
{\it School of Mathematics and Physics, University of Tasmania}\\
{\it GPO Box 252-37, Hobart Tasmania 7001, Australia}\\
{\it September 2005}\\
\end{center}
\par \vskip .3in \noindent

\vspace{1cm} \noindent\textbf{Abstract} \normalfont 
\\\noindent
Distance based algorithms are a common technique in the construction of phylogenetic trees from taxonomic sequence data. The first step in the implementation of these algorithms is the calculation of a pairwise distance matrix to give a measure of the evolutionary change between any pair of the extant taxa. A standard technique is to use the $\log\det$ formula to construct pairwise distances from aligned sequence data. We review a distance measure valid for the most general models, and show how the $\log\det$ formula can be used as an estimator thereof. We then show that the foundation upon which the $\log\det$ formula is constructed can be generalized to produce a previously unknown estimator which improves the consistency of the distance matrices constructed from the $\log\det$ formula. This distance estimator provides a consistent technique for constructing quartets from phylogenetic sequence data under the assumption of the most general Markov model of sequence evolution.

\vfill
\hrule \mbox{} \\
{\footnotesize{ 
$^*$ Australian Postgraduate Award}\\
{$^{\dagger}$ Alexander von Humboldt Fellow}\\
{\textit{keywords:} phylogenetics, distance methods, $\log\det$, tangle}\\
{\textit{email:} Jeremy.Sumner@utas.edu.au, Peter.Jarvis@utas.edu.au}
}
\end{titlepage}
\section{Introduction}

The distance based approach to phylogenetic reconstruction using the neighbor joining algorithm is a commonly used technique \cite{gascuel1997,lake1994,pearson1999,saitou1987}. Under the assumptions of a Markov model of sequence evolution, the phylogenetic relationship is uniquely reconstructible from (suitably defined) pairwise distances \cite{steel1998}. The approach relies crucially upon the calculation of distance matrices from aligned sequence data which give a measure of the pairwise evolutionary distance between the extant taxa under consideration. As far as tree building algorithms are concerned it is required that the distances are strictly \textit{linearly} related to the sum of the (theoretical) edge lengths of the phylogenetic tree, and that the parameters of the linear relation do not vary across the tree. It is essential to the analysis that the measure of distance chosen has both biological and statistical as well as mathematical significance. If one assumes the standard Markov model, the edge lengths of a phylogenetic tree can be taken mathematically to be a quantity which we refer to as the \textit{stochastic distance}. (For mathematical discussion of this quantity see Goodman \cite{goodman1973} who refers to the stochastic distance as \textit{intrinsic time}, and see also Barry and Hartigan \cite{barry1987} who gave a biological interpretation.) Under the assumptions of a general Markov model the $\log\det$ formula is commonly used to obtain pairwise distances. Further, if one may assume a stationary process then the $\log\det$ formula can be modified to give an estimate of the actual stochastic distance \cite{lockhart1994}. (That is, the constants of the linear relation are set by the stationarity assumption.) \\\indent
Distance based methods and, consequently, the $\log\det$ formula are often used in favour of other methods (such as maximum likelihood) in cases where there has been significant compositional heterogeneity during the evolutionary history. The theoretical basis which motivates this usage was presented by Steel \cite{steel1994} and is discussed in Lockhart, Steel, Hendy and Penny \cite{lockhart1994} and Gu and Li \cite{gu1996}. More recently, Jermiin, Ho, Ababneh, Robinson and Larkum published a simulation study which confirms that the $\log\det$ outperforms other techniques in this case \cite{jermiin2004}. Lockhart \textit{et al.} showed that by using the assumption that the base composition remains close to constant, the $\log\det$ formula can be modified to give an estimate of the actual stochastic distance.  However, as will be shown, in both its original and modified form the $\log\det$ formula includes an approximation which is crucially dependent upon the compositional heterogeneity remaining minimal. The effectiveness of the $\log\det$ formula to correctly reconstruct the phylogenetic history when there is been significant compositional heterogeneity is thus brought into question. Hence there is a contradictory state of affairs between the theoretical basis of the $\log\det$ and the circumstances under which it is implemented. In this paper we will generalize the $\log\det$ formula in such a way that this dependence upon base composition is truly absent.\\\indent
A disadvantage of the $\log\det$ formula is that it uses only \textit{pairwise} sequence data and is blind to the fact that extra information regarding pairwise distances can be obtained from the sequence data of additional taxa. Felsenstein \cite{felsenstein2004} mentions that it is surprising that distance techniques work at all given that they ignore the extra information in higher order alignments. This paper details exactly how the $\log\det$ formula can be improved upon by taking functions of aligned sequence data for \textit{three} taxa at a time. It may seem counter-intuitive that consideration of a third taxon can impart information regarding the evolutionary distance between two taxa, but it is the case that by considering a third taxon the $\log\det$ formula can be refined. This result depends crucially upon the fact that, as is somewhat trivially the case for two taxa, there is only one possible (unrooted) tree topology relating three taxa. (For discussion of what a tree topology is see \cite{nei2000}, Chapter 5.) It is possible to refine the $\log\det$ formula by considering the respective distance to an arbitrary third taxon. (The reader should note that the use of triplet sequence data to the problem of reconstruction of the Markov model also was also considered in \cite{chang1996} and \cite{pearl1986}. The approach discussed in the present work is original in the sense that triplets of the aligned sequences are being used explicitly in a distance method, and follows on from the theoretical discussions of \cite{sumner2005}.)
\\\indent
A complication arises regarding the total stochastic distance between leaves and the placement of the root of a phylogenetic tree. It turns out that if we define phylogenetic trees of identical topology to be equivalent if they give the identical probability distributions then we find that the total stochastic distance between leaves is not, in general, left unchanged as we move the root of the tree. The so defined equivalence class provides a generalization of Felsenstein's \textit{pulley principle} \cite{fels1981} and was first presented in Steel, Szekely and Hendy \cite{steel1994b}. The fact that the stochastic distance is not left unchanged is a surprising result and has important implications regarding the interpretation of the edge lengths of phylogenetic trees defined under the Markov model. In particular this result implies that the $\log\det$ technique is an inconsistent estimator of pairwise distances on phylogenetic trees. It is the purpose of this work to present a new estimator which is consistent in the case of phylogenetic quartets. We are motivated to present this construction of quartet distance matrices by the interest in phylogenetic reconstruction of large trees from the correct determination of the set of $\binom{n}{4}$ quartets \cite{bryant2001,strimmer1996}. 

\section{The general Markov model on phylogenetic trees and stochastic distance}

It is standard to model sequence evolution as a stochastic process. The discrete space $\mathcal{K}$ is associated with molecular units which we refer to as  \textit{bases} and we define $n:=|\mathcal{K}|$. For example, in the case of DNA sequences we have $\mathcal{K}=\{A,G,C,T\}$ and $n=4$. We then consider each instance of a base to be a random variable $X\in\mathcal{K}$ and the stochastic time evolution of sequence data is modelled as a continuous time Markov chain (CTMC) such that
\beqn\label{ctmc}
\frac{d}{dt} \mathbb{P}(X(t)=i)=\sum_{j}\mathbb{P}(X(t)=j)q_{ji}(t),\qquad i,j\in\mathcal{K}.
\eqn
The $q_{ij}(t)$ are called \textit{rate parameters} and satisfy the relations
\beqn\label{rates}
q_{ij}(t)\geq 0,\quad \forall i\neq j;\qquad q_{ii}(t)=-\sum_{j\neq i} q_{ij}(t).
\eqn
We define $Q(t)=\left[q_{ij}(t)\right]_{(i,j\in\mathcal{K})}$ as the \textit{rate matrix} associated with the Markov chain. The Markov chain is called \textit{homogeneous} if the rate matrix is time independent. The results presented in this paper are equally valid for inhomogeneous models where the rate matrix is time dependent and so we allow for this generality throughout. It is also common to impose further symmetries upon the rate matrix such as the Jukes Cantor and  Kimura 3ST models \cite{nei2000}. However, the results presented here are again valid for any rate matrix satisfying (\ref{rates}), and hence no restriction upon the rate parameters is made.\\\indent
For notational simplicity we will write $\pi_{i}(t):=\mathbb{P}(X(t)=i)$ and, given an initial distribution $\pi_i(0)$, write solutions of (\ref{ctmc}) as
\beqn
\pi_i(t)=\sum_{j\in\mathcal{K}}\pi_j(s)m_{ji}(s,t),\qquad 0\leq s< t;\nonumber
\eqn
where $m_{ij}(s,t):=\mathbb{P}(X(t)=j|X(s)=i)$ are the \textit{transition probabilities} of the chain. We define the matrix $M(s,t)=\left[m_{ij}(s,t)\right]_{(i,j\in\mathcal{K})}$ such that in the homogeneous case the transition probabilities only depend on the difference $(t-s)$ and can be represented in terms of the rate matrix as
\beqn
M(s,t)=M(0,t-s)=e^{Q[(t-s)]}:=\sum_{n=0}^{\infty}\frac{Q^n[(t-s)]^n}{n!}.\nonumber
\eqn
In the inhomogeneous case there are several representations available for the matrix of transition probabilities (for details see \cite{isoifescu1980,rindos1995}). The representation that is of most use to us here is the time ordered product, which can be written for sufficiently small $\delta t$ in the approximate form
\beqn\label{inhom}
M(s,t)&\simeq M(s,s+\delta t)M(s+\delta t,s+2\delta t)..M(t-2\delta t,t-\delta t)M(t-\delta t,t)\\
&=e^{Q(s)\delta t}e^{Q(s+\delta t)\delta t}..e^{Q(t-2\delta t)\delta t}e^{Q(t-\delta t)\delta t}.
\eqn
From this solution it is easy to show that the \textit{backward} and \textit{forward} Kolmogorov equations: 
\beqn\label{kalamorov}
\frac{\partial M(s,t)}{\partial s}&=-Q(s)M(s,t),\\
\frac{\partial M(s,t)}{\partial t}&=M(s,t)Q(t),
\eqn
are satisfied as required of any CTMC \cite{isoifescu1980}.

\subsection{Stochastic distance}

In this work we will be interested in the assignment of edge lengths to phylogenetic trees. To this end we consider the rate of change of base changes at time $s$: \footnote{It is standard to include a factor of $n^{-1}$ in this definition. However, this factor clutters the consequent formulae and here we do not include it as it has no consequence to the forgoing discussion and can always be incorporated into the analysis later.}
\beqn
\lambda(s):=\sum_{i\in\mathcal{K}}\frac{\partial\mathbb{P}(X(t)=i|X(s)\neq i)}{\partial t}|_{t=s}.\nonumber
\eqn
By considering (\ref{rates}) and (\ref{kalamorov}) this quantity can be explicitly expressed using the rate parameters: 
\beqn
\lambda(s)&=-\sum_{i}q_{ii}(s),\\\nonumber
&=-trQ(s).
\eqn
From these considerations we define the \textit{stochastic distance} to be given by the expression
\beqn
\omega(s,t):=\int_s^t\lambda(u)du.\nonumber
\eqn
By considering the time ordered product representation (\ref{inhom}) and the Jacobi identity $\det e^X=e^{trX}$, we find that the stochastic distance can be directly related to the transition probabilities of the Markov chain:
\beqn\label{omega}
\omega(s,t)=-\ln\det M(s,t).
\eqn
Our assignment of edge lengths will take the Markov matrix associated with each edge and set the edge length equal to the stochastic distance. 
\\\indent
The relation (\ref{omega}) is known in various guises in both the mathematical and phylogenetic literature \cite{barry1987,goodman1973} and, as will be confirmed in the next section, is the basis of the $\log\det$ formula. It should also be noted that (\ref{omega}) will remain positive and finite because $\omega(s,s)=0$, $\lambda(s)\geq 0$ and the integral $\int_0^T \lambda(t)dt$ is not expected to diverge.\footnote{There are two cases where the integral may diverge, but we can safely exclude these possibilities as follows. i. $\lambda(t)$ may be a badly behaved function. We can reject this possibility outright in phylogenetics as there is every reason to expect the rate parameters to change smoothly with time. ii. $T\rightarrow\infty$. We can safely ignore this possibility as we will be assuming that the divergence times of the Markov chain are sufficiently small such that the phylogenetic historical signal is still obtainable.}

\subsection{Phylogenetic trees}

\begin{figure}[t]
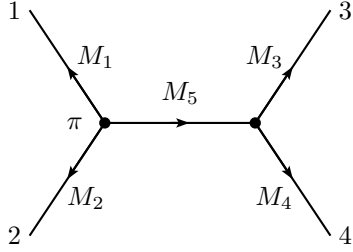

\centering
\pspicture[](3,)(,)
\psset{linewidth=\pstlw,xunit=0.5,yunit=0.5,runit=0.5}
\psset{arrowsize=2pt 2,arrowinset=0.2}
\psline{-}(2,3)(0,0)
\rput(-0.4,0){2}
\psline{-}(2,3)(0,6)
\rput(-0.4,6){1}
\psline{-}(2,3)(6,3)
\psline{-}(6,3)(8,6)
\rput(8.4,6){3}
\psline{-}(6,3)(8,0)
\rput(8.4,0){4}
\psline{->}(2,3)(1,4.5)
\rput(1.75,4.75){$M_1$}
\psline{->}(2,3)(1,1.5)
\rput(1.5,1.0){$M_2$}
\psline{->}(6,3)(7,4.5)
\rput(6.25,4.75){$M_3$}
\psline{->}(6,3)(7,1.5)
\rput(6.5,1.0){$M_4$}
\psline{->}(3.75,3)(4.25,3)
\rput(4,3.8){$M_5$}
\pscircle[linewidth=0.8pt,fillstyle=solid,fillcolor=black](2,3){.15} 
\pscircle[linewidth=0.8pt,fillstyle=solid,fillcolor=black](6,3 ){.15} 
\rput(1.2,3){$\pi $}
\endpspicture 
\caption{Phylogenetic tree of four taxa}
\label{pic:fourleaf}
\end{figure}

The remaining task is to model the case of multiple taxa evolving under a stochastic process. Effectively the model consists of multiple copies of the random variable $X(t)$ taken as a generalization (via a tree structure) of a cartesian product and then modelled collectively as a CTMC. The reader is referred to \cite{semple2003} for a more extended discussion of the model. Here we keep the presentation to a minimum while allowing for the introducing of some essential notation and concepts.
\\\indent 
A tree is a connected graph without cycles and consists of a set of vertices and edges $T=(V,E)$. Vertices of degree one are called \textit{leaves} and we partition the set of vertices as $V=L\cup N$ where $L$ is the set of leaves and $N$ is the set of internal vertices. We direct each edge of $T$ away from a distinguished vertex, $\pi$, known as the \textit{root} of the tree. Consequently, a given edge lying between vertices $u$ and $v$ is specified as an ordered pair $e=(u,v)$, where $u$ lies on the (unique) path between $v$ and $\pi$. The stochastic phylogenetic model is then made by assigning a set of random variables $\{X_e,\ e\in E\}$ to each edge of the tree; these random variables are assumed to be conditionally independent and individually satisfy the properties of a CTMC. Taking a distribution at the root of the tree, $\{\mathbb{P}(X_\pi=i):=\pi_i, i\in\mathcal{K}\}$, completes the phylogenetic tree. The interpretation of a phylogenetic tree is that the probability distribution at each leaf is associated with the observed sequence of a single taxon and the joint probability distribution across a number of leaves is associated with the aligned sequences of the same number of taxa.   
\\\indent
For example in Figure \ref{pic:fourleaf} we present the tree consisting of 4 taxa which has probability distribution
\beqn
p_{i_1i_2i_3i_4}=\sum_{j,k}\pi_jm^{(5)}_{jk}m^{(1)}_{ji_1}m^{(2)}_{ji_2}m^{(3)}_{ki_3}m^{(4)}_{ki_4},\nonumber
\eqn
where 
\beqn
p_{i_1i_2i_3i_4}:=\mathbb{P}(X_1=i_1,X_2=i_2,X_3=i_3,X_4=i_4)\nonumber
\eqn
and we refer to these quantities as \textit{pattern probabilities}.

\section{Pairwise distance measures}\label{pairwise}

In this section we will derive and discuss a standard approach to the construction of distance matrices. (For an excellent perspective of the various measures of phylogenetic pairwise distance see \cite{baake1999}.) A distance matrix, $\phi=[\phi_{ab}]_{(a,b)\in L}$, is constructed from the aligned sequence data of multiple extant taxa such that each entry gives a suitable estimate of the distance between a given pair of taxa. The mathematical conditions on the $\phi_{ab}$ are the standard conditions of a distance function as well as the four point condition \cite{steel1998} (which is required for the distance measure to be consistent with the tree structure):
\beqn\label{distance}
\phi_{ab}&\geq 0,\\
\phi_{ab}&=0\text{ iff }a=b,\\
\phi_{ab}&=\phi_{ba},\\
\phi_{ab}+\phi_{cd}&\leq\textit{max}\{\phi_{ac}+\phi_{bd},\phi_{ad}+\phi_{bc}\};\qquad\forall\ a,b,c,d\in L.
\eqn
There are no further conditions required upon $\phi$ for it to give a unique tree reconstruction \cite{steel1998}. However it is of course desirable for the distance measure to have a well defined biological interpretation. To this end, for a given edge $e$, we define the \textit{edge length}, $\omega_e$, which we set to be the stochastic distance (\ref{omega}) taken from the Markov model:
\beqn
\omega_e=-\ln\det M_e.\nonumber
\eqn
It is then apparent that any significant estimate of pairwise distance must statistically be expected to converge to a value which is linearly related to the sum of the stochastic distances lying on the (unique) path between the two taxa under consideration. It should be clear that such a measure will satisfy the relations (\ref{distance}). It is crucial to the performance of the distance measure under a tree building algorithm that the parameters of the linear relation are expected to be \textit{constant} for all pairs of taxa. That is, given the unique path between leaf $a$ and $b$, $P(T;a,b)$, we are demanding that statistically we have the following convergence:
\beqn
\phi_{ab}\rightarrow\alpha\omega{(a,b)}+\beta,\nonumber
\eqn
where
\beqn
\omega{(a,b)}:=\sum_{e\in P(T;a,b)}\omega_e,\nonumber
\eqn
and $\alpha$ and $\beta$ are expected to be independent of $a$ and $b$. As we will see, the $\log\det$ formula does not satisfy this property for the most general models.
\\\indent
\subsection{The $\log\det$ formula}
\begin{figure}[t]
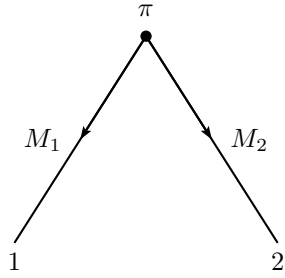

\centering
\pspicture[](3,)(,)
\psset{linewidth=\pstlw,xunit=0.5,yunit=0.5,runit=0.5}
\psset{arrowsize=2pt 2,arrowinset=0.2}
\psline{->}(4,6)(2.25,3.25)
\psline{->}(4,6)(5.75,3.25)
\psline{-}(4,6)(0.5,0.5)
\psline{-}(4,6)(7.5,0.5)
\pscircle[linewidth=0.8pt,fillstyle=solid,fillcolor=black](4,6){0.15} 
\rput(4,6.7){$\pi$}
\rput(1.25,3.25){$M_1$}
\rput(6.75,3.25){$M_2$}
\rput(0.5,0){1}
\rput(7.5,0){2}
\endpspicture 
\caption{Phylogenetic tree of two taxa.}
\label{pic:twotaxa}
\end{figure}

\begin{figure}[t]
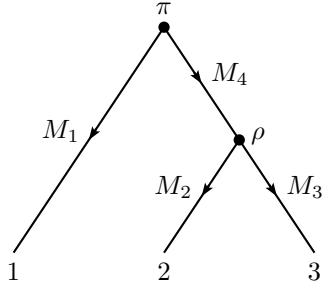

\centering
\pspicture[](3,)(,)
\psset{linewidth=\pstlw,xunit=0.5,yunit=0.5,runit=0.5}
\psset{arrowsize=2pt 2,arrowinset=0.2}
\psline{->}(4,6)(2,3)
\rput(1.25,3.25){$M_1$}
\pscircle[linewidth=0.8pt,fillstyle=solid,fillcolor=black](4,6){0.15}
\rput(4,6.5){$\pi$} 
\psline{-}(2.2,3.3)(0,0)
\rput(0,-.5){1}
\psline{->}(4,6)(5,4.5)
\rput(5.75,4.75){$M_4$}
\psline{-}(4.8,4.8)(6,3)
\pscircle[linewidth=0.8pt,fillstyle=solid,fillcolor=black](6,3){0.15}
\rput(6.5,3.1){$\rho$}
\psline{->}(6,3)(5,1.5)
\rput(4.25,1.75){$M_2$}
\psline{-}(5.2,1.8)(4,0)
\rput(4,-.5){2}
\psline{->}(6,3)(7,1.5)
\rput(7.75,1.75){$M_3$}
\psline{-}(6.8,1.8)(8,0)
\rput(8,-.5){3}
\endpspicture  
\caption{Phylogenetic tree of three taxa}
\label{pic:threetaxaroot}
\end{figure}

In Figure \ref{pic:twotaxa} we consider the two taxa phylogenetic tree, with pattern probabilities given by
\beqn\label{twotaxa}
p_{i_1i_2}=\sum_{j}\pi_jm^{(1)}_{ji_1}m^{(2)}_{ji_2}.
\eqn
By considering the matrices defined as 
\beqn
P^{(1,2)}:&=\left[p_{ij}\right]_{(i,j)\in\mathcal{K}},\\
D_\pi:&=\left[diag(\pi_i)\right]_{i\in\mathcal{K}};\nonumber
\eqn
 it is easy to show that (\ref{twotaxa}) is equivalent to
\beqn
P^{(1,2)}=M_1D_\pi M_2^t.\nonumber
\eqn
Taking the determinant of this expression and considering (\ref{omega}) yields
\beqn\label{conc}
\det P^{(1,2)} &=\det M_1\det M_2\det D_\pi\\
&=e^{-(\omega_1+\omega_2)}\prod_i\pi_i.
\eqn
This expression can be generalized to the case of any two taxa from a given phylogenetic tree:
\beqn\label{det}
\det P^{(a,b)}=e^{-\omega(a,b)}\prod_i\pi_i^{(a,b)},
\eqn
where $\pi_i^{(a,b)}$ is the distribution at the most recent ancestral vertex between taxa $a$ and $b$ determined by the meeting point of the two paths traced backwards along the phylogenetic tree from leaf $a$ and $b$.\\\indent
Now $\omega(a,b)$ is theoretically equal to the total stochastic distance between each of $a$ and $b$ and their most recent ancestral vertex and hence it is clear that $-\log\det P^{(a,b)}$ will be linearly related to this quantity. In the original formulation of the $\log\det$, a distance measure between two taxa was defined as
\beqn\label{logdet}
d_{ab}:&=-\log\det P^{(a,b)}\\
&=\omega(a,b)-\sum_i\ln[\pi_i^{(a,b)}],
\eqn
and shown to satisfy the conditions (\ref{distance}) \cite{steel1994}. From this relation it seems that one can take $\alpha=1$ and $\beta=-\sum_i\ln[\pi_i^{(a,b)}]$  and evaluate (\ref{logdet}) on the observed pattern frequencies for each pair of taxa to calculate a well defined distance matrix from a set of aligned sequence data (as was presented in \cite{lockhart1994}). This procedure depends crucially upon the shifting term $\beta=\sum_i\ln[\pi_i^{(a,b)}]$ being independent of $a$ and $b$. However, this is only true in special circumstances such as star phylogeny or if the base composition is constant (the stationary model). In the general case, one is led to a different shifting term depending on the topology of the tree (this was noted in Sumner and Jarvis \cite{sumner2005} and we reproduce the result here). Consider the phylogenetic tree of three taxa given in Figure \ref{pic:threetaxaroot}
with pattern probabilities given by
\beqn\label{threetaxa}
p_{i_1i_2i_3}=\sum_{j,k}\pi_jm^{(1)}_{ji_1}m^{(4)}_{jk}m^{(2)}_{ki_2}m^{(3)}_{ki_3}.\nonumber
\eqn
By calculating (\ref{logdet}) for the three possible pairs of taxa we find that
\beqn
d_{12}&=(\omega_1+\omega_4+\omega_2)-\sum_i\ln\pi_i,\\\nonumber
d_{13}&=(\omega_1+\omega_4+\omega_3)-\sum_i\ln\pi_i,\\
d_{23}&=(\omega_1+\omega_3)-\sum_i\ln\rho_i,
\eqn
from which it is explicitly clear that the shifting term is \textit{not} constant across this phylogenetic tree. The shifting term is dependent on the base composition at the most recent ancestral node of the two taxa and from the above example it is clear that this depends on the topology of the tree and is not always simply the root of the tree. This means that (\ref{logdet}) does not produce distance matrices whose entries are linearly related to the edge length of the tree because the entries of the matrix will depend essentially upon the topology of the tree. 
\\\indent
It is, however, possible to obtain an estimate of the total stochastic distance between any two taxa by modifying the $\log\det$ formula. The ancestral base composition is approximated by using the harmonic mean
\beqn\label{harmonicmean}
\prod_i\pi^{(a,b)}_i\approx[\prod_{i_1,i_2}\pi^{(a)}_{i_1}\pi^{(b)}_{i_2}]^{\frac{1}{2}},
\eqn
where $\pi^{(a,b)}_k$ is the closest common ancestral base composition between taxa $a$ and $b$ and $\pi^{(a)}_i:=\mathbb{P}(X_a(\tau_a)=i)$ (and similarly for $b$). One is then led to the formula
\beqn\label{logdetharm}
d'_{ab}:=-\ln\det P^{(a,b)}+\fra{1}{2}\sum_{i_1,i_2}(\ln\pi^{(a)}_{i_1}+\ln\pi^{(b)}_{i_2}),\qquad\forall\ a,b\in L.
\eqn
where $d'_{ab}$ is then an estimator of the total stochastic distance between taxa $a$ and $b$. (This form of the $\log\det$ formula was presented in \cite{lockhart1994} and \cite{steel1998}).
\\\indent
In the case of a stationary base composition model the additional assumption is made that
\beqn
\sum_{j}\pi_jm^{(e)}_{ji}=\pi_i;\qquad\forall\ e\in E.\nonumber
\eqn
In this case we have
\beqn
\pi_i^{(a,b)}=\pi^{(a)}_i=\pi^{(b)}_i;\qquad\forall\ a,b\in L.\nonumber
\eqn
and it is clear that the the harmonic mean approximation becomes an exact relation and the $\log\det$ formula is expected to converge exactly to the total stochastic distance between the two taxa.

\subsection{The tangle}

In this section we will show how the $\log\det$ formula can be generalized to obtain, for the most general Markov models,  an unbiased estimate of the distance matrix. The basis of the technique is the existence a measure analogous to (\ref{conc}) which is valid for \textit{triplets}. \\\indent
Sumner and Jarvis \cite{sumner2005} presented a polynomial function $\mathcal{T}$ which is known in quantum physics as the \textit{tangle} and can be evaluated on phylogenetic data sets of three aligned sequences in the case of $n=2$. Evaluated on the pattern probabilities of any phylogenetic tree of three taxa, $\{a,b,c\}$, the tangle takes on the theoretical value 
\beqn\label{tangle}
\mathcal{T}(a,b,c)=e^{-\omega(a,b,c)}\left(\prod_{i\in\mathcal{K}}\pi_i\right)^2,
\eqn
where 
\beqn
\omega{(a,b,c)}:=\sum_{e\in T}\omega_e\nonumber
\eqn
$\pi$ is the common ancestral root of the three taxa and this relation holds independently of the particular tree topology which relates $\{a,b,c\}$. This independence upon the topology is a very nice property and is crucial to the practical use of the tangle as a distance measure. The similarity between (\ref{tangle}) and (\ref{det}) should be noted.\\\indent
In this work we report generalized tangles, which are polynomials which satisfy (\ref{tangle}) for the cases of $n=3,4$ in addition to the $n=2$ case which was presented in \cite{sumner2005}. It is possible to infer the existence of the tangles and derive their polynomial form from group theoretical considerations. Here we give forms using the the completely antisymmetric (Levi-Civita) tensor, $\epsilon$, which has components $\epsilon_{i_1i_2...i_n}$ and satisfies $\epsilon_{12...n}=1$. For the cases of $n=2,3,4$ the tangles are given by\footnote{This expression for $\mathcal{T}_2$ corrects for the erroneous expression presented in \cite{sumner2005}.}

\noindent
$
\mathcal{T}_2=\fra{1}{2!}\sum_1^2 p_{i_1i_2i_3}p_{j_1j_2j_3}p_{k_1k_2k_3}p_{l_1l_2l_3}\epsilon_{i_1j_1}\epsilon_{i_2j_2}\epsilon_{k_1l_1}\epsilon_{k_2l_2}\epsilon_{i_3l_3}\epsilon_{j_3k_3},$

\noindent
$
\mathcal{T}_3=\fra{1}{3!}\sum_1^3p_{i_1i_2i_3}p_{j_1j_2j_3}p_{k_1k_2k_3}p_{l_1l_2l_3}p_{m_1m_2m_3}p_{n_1n_2n_3}\epsilon_{i_1j_1k_1}\epsilon_{j_2k_2l_2}\epsilon_{k_3l_3m_3}\epsilon_{l_1m_1n_1}\epsilon_{m_2n_2i_2}\epsilon_{i_3j_3k_3},$

\noindent
$
\mathcal{T}_4=\fra{1}{4!}\sum_1^4p_{i_1j_1k_1}p_{i_2j_2k_2}p_{i_3j_3k_3}p_{i_4j_4k_4}p_{i_5j_5k_5}p_{i_6j_6k_6}p_{i_7j_7k_7}p_{i_8j_8k_8}$\\ 
\hspace*{140pt}$\cdot\epsilon_{i_1i_2i_3i_4}\epsilon_{i_5i_6i_7i_8}\epsilon_{j_1j_5j_4j_8}\epsilon_{j_2j_6j_3j_7}\epsilon_{k_1k_5k_2k_6}\epsilon_{k_3k_7k_4k_8};
$

\noindent
respectively, (where the summation is over every index). The expression (\ref{tangle}) can be proved by studying the group theoretical properties of the tangle (see \cite{sumner2005}) and by explicitly expanding the above forms. For the tangle on two characters we find
\beqn
\mathcal{T}_2=&-p_{122}^2p_{211}^2 + 2p_{121}p_{122}p_{211}p_{212} - p_{121}^2p_{212}^2 + 2p_{112}p_{122}p_{211}p_{221} + 2p_{112}p_{121}p_{212}p_{221} - \nonumber
\\&4p_{111}p_{122}p_{212}p_{221} - p_{112}^2p_{221}^2 - 4p_{112}p_{121}p_{211}p_{222} + 
2p_{111}p_{122}p_{211}p_{222} 
\\&+ 2p_{111}p_{121}p_{212}p_{222} + 
2p_{111}p_{112}p_{221}p_{222} - p_{111}^2p_{222}^2.
\eqn
Substantial computer power is required to explicitly compute $\mathcal{T}_3$ and $\mathcal{T}_4$. These polynomials have 1152 and 431424 terms, respectively. The expansions has been achieved by the authors, who can be contacted to obtain a practical algorithm, or the polynomials themselves.

\subsection{Star topology}
Consider the phylogenetic tree relating three taxa with a star topology:
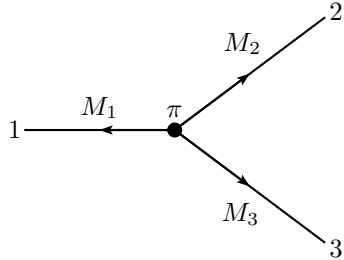
\beqn
\\
\\
\pspicture[0.5](5,0)(1,2.5)
\psset{linewidth=\pstlw,xunit=0.5,yunit=0.5,runit=0.5}
\psset{arrowsize=2pt 2,arrowinset=0.2}
\psline{-}(4,3)(0,3)
\rput(-.25,3){1}
\psline{-}(4,3)(8,0)
\rput(8.3,-0.15){3}
\psline{-}(4,3)(8,6)
\rput(8.3,6.15){2}
\psline{->}(4,3)(2,3)
\rput(2,3.6){$M_1$}
\psline{->}(4,3)(6,4.5)
\rput(5.8,5.3){$M_2$}
\psline{->}(4,3)(6,1.5)
\rput(5.75,.75){$M_3$}
\pscircle[linewidth=0.8pt,fillstyle=solid,fillcolor=black](4,3){.2} 
\rput(4,3.5){$\pi$}
\endpspicture  \nonumber
\\
\\
\eqn
with pattern probabilities given by the formula
\beqn\label{threefreq}
p_{i_1i_2i_3}=\sum_{j}\pi_jm^{(1)}_{ji_1}m^{(2)}_{ji_2}m^{(3)}_{ji_3}.
\eqn
 Here we will use the fact that the root of this tree is also the common ancestral root of any pair of the three taxa. (This is not the case in general if we allow for a general rooting of the tree and/or more than three taxa. The complications arising in these cases will be dealt with in the next section.)
 \\\indent
Considering the formulae (\ref{tangle}) and (\ref{conc}) we are led to introduce the novel distance matrix, $\Delta$, with the pairwise distance between $\{a,b\}$ given by
\beqn\label{dist}
\Delta^{(c)}_{ab}:=-\ln\mathcal{T}(a,b,c)+\ln{\det{P^{(a,c)}}+\ln\det{P}^{(b,c)}},\qquad a,b,c\in L.
\eqn
From (\ref{conc}) and (\ref{tangle}) it follows that 
\beqn
\Delta^{(c)}_{ab}=\omega(a,b),\nonumber
\eqn
such that our new formula will directly give the stochastic distance between the two taxa. There is no need to make the harmonic mean approximation and this distance measure is mathematically and biologically meaningful. This is the main result of this paper: given a set of aligned sequence data, the tangle formula (\ref{dist}) can be used to compute the \textit{exact} pairwise edge lengths for any triplet. As mentioned above, the explicit polynomial form of the tangle has been computed for the cases of two, three and four bases and it is our intent that (\ref{dist}) will provide a significant improvement over the $\log\det$ formula in the calculation of pairwise distance matrices for these cases.

\subsection{Summary}
Considering the stochastic distance to be the correct way to assign edge lengths to branches of a phylogenetic tree, we have reviewed three different ways of obtaining a distance measure between any two taxa $a$ and $b$:
\begin{enumerate}
\item {$d_{ab}=-\ln\det P^{(a,b)}$}
\item {$d'_{ab}=-\ln\det P^{(a,b)}+\frac{1}{2}\sum_{i_1,i_2}(\ln\pi^{(a)}_{i_1}+\ln\pi^{(b)}_{i_2})$}
\item {$\Delta^{(c)}_{ab}=-\ln\mathcal{T}(a,b,c)+\ln{\det{P^{(a,c)}}+\ln\det{P}^{(b,c)}}$}
\end{enumerate}
where one substitutes the observed pattern frequencies into these expressions. From the previous considerations we found that these three distance measures have the following properties:
\begin{enumerate}
\item When $d_{ab}$ is evaluated on a set of observed pattern frequencies, this estimator satisfies the requirements of a distance function (\ref{distance}), but is inconsistent with the general Markov model as the estimate is \textit{not} expected to converge to a value that is linearly related to $\omega(a,b)$.
\item When $d'_{ab}$ is evaluated on a set of observed pattern frequencies, this estimator satisfies the requirements (\ref{distance}) and is expected to converge to a value that is linearly related to $\omega(a,b)$ whenever the compositional heterogeneity is absent. In the heterogeneous case this quantity approximates $\omega(a,b)$ by using (\ref{harmonicmean}).
\item When $\Delta^{(c)}_{ab}$ is evaluated on a set of observed pattern frequencies, this estimator satisfies the requirements of (\ref{distance}) and is expected to converge \text{exactly} to $\omega(a,b)$ in all cases.
\end{enumerate}
\indent
Thus we see that the tangle formula (\ref{dist}) should be a significant improvement as an empirical estimator of $\omega(a,b)$ upon both forms of the $\log\det$ formula. However, the formula (\ref{dist}) depends on taking an arbitrary third taxon, $c$. The question remains as to what to do in the case of constructing pairwise distances for sets of greater that three taxa. The surprising answer to this question will be addressed in the next section where we will bring into question the uniqueness of the theoretical quantity $\omega(a,b)$. The discussion has consequences for the interpretation of each of the estimators of pairwise distances that we have discussed.
\\\indent

\section{Generalized pulley principle}

In this section we generalize the Felsenstein's pulley principle \cite{fels1981}. In its original formulation the pulley principle describes the unrootedness of phylogenetic trees where the underlying Markov model is assumed to be reversible and stationary. Here we show how the pulley principle may be generalized to remain valid under the most general Markov models. Our immediate motivation is to show that (\ref{dist}) remains a valid distance measure under the circumstance of a general phylogenetic tree of multiple taxa. Unfortunately this generalization introduces surprising mathematical complications which have consequences not only for our formula (\ref{dist}), but also for the $\log\det$ technique and any other estimate of the stochastic distance upon a phylogenetic tree. The discussion will lead to the consequence that, for a given tree topology, there are multiple -- actually, infinitely many -- phylogenetic trees with identical probability distributions. (These phylogenetic trees differ by arbitrary rerootings and consequential re-direction of edges.)  We will see that the generalized pulley principle shows that as far as inference from the observed pattern frequencies is concerned, there is no theoretical justification behind specifying the root of a phylogenetic tree if the most general Markov model is allowed. Also, we will see that the theoretical value of the stochastic distance is not constant for arbitrary rerootings of a phylogenetic tree. Clearly, if the stochastic distance is not uniquely defined theoretically, then one must be careful in interpreting any formula which gives an estimate thereof from the observed data.\\\indent
Considering a phylogenetic tree as a directed graph shows that a rerooting involves redirecting an edge (or part thereof). The property required is that the Markov chain on the involved edge is taken to progress as if time has been reversed, and we refer to the new chain as the \textit{time-reversed} chain. This should be compared to the requirement of \textit{reversibility} as defined in the mathematical literature, (for example see \cite{isoifescu1980}). In the case of a stationary and reversible Markov chain the time-reversed chain (as we will define) is identical to the original chain.
\\\indent 
By way of example, we take the rooted tree of three taxa (\ref{threetaxa}) and redirect the relevant internal edge to give the following rerooting:
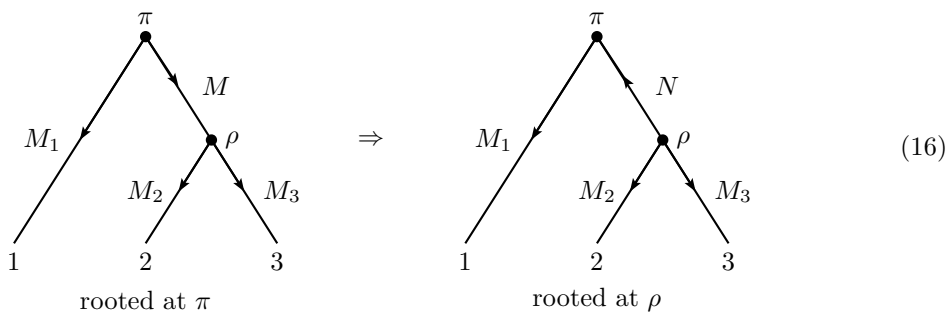
\beqn
\\
\\
\pspicture[0.5](8.75,0)(1,2.5)
\psset{linewidth=\pstlw,xunit=0.5,yunit=0.5,runit=0.5}
\psset{arrowsize=2pt 2,arrowinset=0.2}
\psline{->}(4,6)(2.25,3.25)
\psline{->}(4,6)(4.875,4.675)
\psline{-}(4,6)(5.75,3.25)
\psline{-}(4,6)(0.50,0.50)
\psline{->}(5.75,3.25)(6.625,1.875)
\psline{-}(5.75,3.25)(7.50,0.50)
\psline{->}(5.75,3.25)(4.875,1.875)
\psline{-}(5.75,3.25)(4,0.5)
\pscircle[linewidth=0.8pt,fillstyle=solid,fillcolor=black](4,6){.15}
\pscircle[linewidth=0.8pt,fillstyle=solid,fillcolor=black](5.75,3.25){.15}
\rput(4,6.5){$\pi$}
\rput(1.25,3.25){$M_1$}
\rput(5.875,4.675){$M$}
\rput(4,1.875){$M_2$}
\rput(6.3,3.25){$\rho$}
\rput(7.625,1.875){$M_3$}
\rput(0.5,0){1}
\rput(4,0){2}
\rput(7.5,0){3}
\rput(4,-1){\text{rooted at $\pi$}}
\rput(16,-1){\text{rooted at $\rho$}}
\rput(10,3.25){$\Rightarrow$}
\psline{->}(16,6)(14.25,3.25)
\psline{-<}(16,6)(16.875,4.675)
\psline{-}(16,6)(17.75,3.25)
\psline{-}(16,6)(12.50,0.50)
\psline{->}(17.75,3.25)(18.625,1.875)
\psline{-}(17.75,3.25)(19.50,0.50)
\psline{->}(17.75,3.25)(16.875,1.875)
\psline{-}(17.75,3.25)(16,0.5)
\pscircle[linewidth=0.8pt,fillstyle=solid,fillcolor=black](16,6){.15}
\pscircle[linewidth=0.8pt,fillstyle=solid,fillcolor=black](17.75,3.25){.15}
\rput(16,6.5){$\pi$}
\rput(13.25,3.25){$M_1$}
\rput(17.875,4.675){$N$}
\rput(16,1.875){$M_2$}
\rput(18.3,3.25){$\rho$}
\rput(19.625,1.875){$M_3$}
\rput(12.5,0){1}
\rput(16,0){2}
\rput(19.5,0){3}
\endpspicture  
\label{pic:reroot}
\\
\\
\eqn
Our immediate task is to infer the existence of an appropriate time-reversed Markov chain, $N$, such that these two phylogenetic trees give identical probability distributions. If we equate the pattern probabilities of (\ref{pic:reroot}) and contract all edges except the one we are reversing, we are led to the simple algebraic solution
\beqn\label{reversed}
n_{ij}=\frac{m_{ji}\pi_i}{\rho_j}.
\eqn 
(This solution was presented in \cite{steel1994b}.) Presently we use this result to give an explicit form in the general case.
\\\indent
Given a CTMC $X(t)$ with transition probabilities 
\beqn
m_{ij}(s,t):=\mathbb{P}(X(t)=i|X(s)=j),\nonumber
\eqn
we wish to find a second CTMC, $Y(t)$, such that, given any $T\geq 0$, we have
\beqn
\mathbb{P}(Y(t)=i)=\pi_i(T-t),\qquad\forall\quad 0\leq t\leq T.\nonumber
\eqn
That is, if the direction of time is reversed, the second CTMC $Y(t)$ has identical distribution to $X(t)$. The uniqueness of $Y(t)$ is a technical matter which we do not consider, because in the phylogenetic case there are extra restrictions which led to the unique solution (\ref{reversed}).\\\indent 
Considering again the general case, we write
\beqn
\mathbb{P}(Y(t)=j|Y(s)=i):=n_{ij}(s,t)\nonumber
\eqn
and use (\ref{reversed}) to infer the general solution
\beqn\label{reverse}
n_{ij}(s,t)=\frac{m_{ji}(T-t,T-s)\pi_{i}(T-t)}{\pi_{j}(T-s)}.\nonumber
\eqn
It is trivial to show that these transition probabilities satisfy the requirements of a CTMC:
\beqn
\sum_jn_{ij}(s,t)&=1,\qquad\forall\ j,\nonumber\\
N(s,t)N(u,s)&=N(u,t),
\eqn
where $N(s,t)=[n_{ij}(s,t)]_{(i,j\in\mathcal{K})}$.\\\indent
Furthermore, by using (\ref{kalamorov}) we find that the rate parameters of the time-reversed chain can be expressed as
\beqn
f_{ij}(s):&=\frac{\partial n_{ij}(s,t)}{\partial t}|_{t=s}\\
&=\frac{q_{ji}(T-s)\pi_{i}(T-s)}{\pi_{j}(T-s)}-\sum_k\frac{\delta_{ij}q_{ik}(T-s)\pi_{k}(T-s)}{\pi_{j}(T-s)}\nonumber
\eqn
From which it follows that
\beqn
f_{ij}(s)\geq 0,\quad \forall i\neq j;\qquad f_{ii}(s)=-\sum_{j\neq i} f_{ij}(s)\nonumber
\eqn
which confirms that the $f_{ij}(s)$ are a valid set of rate parameters for a CTMC (as expected). It should be noted that even in the case where $X(t)$ is a homogeneous chain it is certainly not the case in general that $Y(t)$ is also homogeneous. Consider, however, the stationary and reversible case, with the respective conditions:
\beqn
\sum_j\pi_i(0)q_{ji}&=0,\\
q_{ij}\pi_j(0)&=q_{ji}\pi_i(0).\nonumber
\eqn
where the stationarity condition ensures that 
\beqn
\pi_i(t)=\pi_i(0),\qquad\forall t.\nonumber
\eqn
 In this circumstance it follows that
\beqn
f_{ij}=q_{ij},\nonumber
\eqn
such that $Y(t)\equiv X(t)$ and is hence also stationary and reversible. This was the basis of Felsenstein's initial formulation of the pulley principle -- if one considers only stationary and reversible Markov chains on a phylogenetic tree, any time-reversed chain is identical to the original Markov chain and hence a phylogenetic tree can be arbitrarily rerooted. We have given a continuous time generalization of Felsenstein's result which removes the stationary and reversible restriction.
\\\indent
Equipped with the solution (\ref{reverse}) it is possible to take any phylogenetic tree and find an alternative tree of identical topology, but rooted in a different place, such that the alternative tree generates an identical probability distribution to that of the original. This is the basis of our generalized pulley principle. 
\\\indent
The reader should note that we have proven that, under the assumptions of the most general Markov model, it is not possible to determine the root of a phylogenetic tree by only considering the probability distribution it generates. Thus, any procedure which determines the root from the observed pattern frequencies must be justified by making additional assumptions about the underlying stochastic process. 
\\\indent
The curious aspect of the general pulley principle is that the stochastic distance is \textit{not} conserved along the edge of the tree where the directedness was reversed. This is easy to show by considering the determinant of (\ref{reverse})
\beqn\label{consistency}
\det N(s,t)=\det M(T-t,T-s)\prod_i\frac{\pi_i(T-t)}{\pi_i(T-s)}
\eqn
Thus the stochastic distance in the reversed time chain is equal to that of the original chain if and only if
\beqn\label{iff}
\prod_i\frac{\pi_i(T-t)}{\pi_i(T-s)}=1.
\eqn
This property of CTMCs and their time-reversed counterparts was observed by Barry and Hartigan \cite{barry1987}. It can be seen that in the stationary case (\ref{iff}) will certainly be true. There are other cases where (\ref{iff}) may hold but there does not seem to any biologically sound way to interpret the required condition. In the proceeding discussion we will consider the consequences of the generalized pulley principle upon the interpretation of distance matrices. We see that for a given observed distribution we can use the generalized pulley principle to show that there are multiple edge length assignments using the stochastic distance which are consistent with the Markov model on a phylogenetic tree. These edge length assignments differ from one another as a consequence of (\ref{consistency}).

\subsection{Interpretation}

\begin{figure}[t]
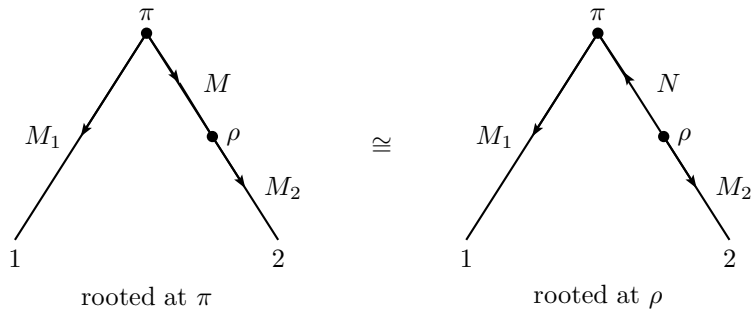

\pspicture[](-2,-1)(0,1)
\psset{linewidth=\pstlw,xunit=0.5,yunit=0.5,runit=0.5}
\psset{arrowsize=2pt 2,arrowinset=0.2}
\psline{->}(4,6)(2.25,3.25)
\psline{->}(4,6)(4.875,4.675)
\psline{-}(4.875,4.675)(5.75,3.25)
\psline{-}(4,6)(0.50,0.50)
\psline{->}(5.75,3.25)(6.625,1.875)
\psline{-}(4,6)(7.50,0.50)
\pscircle[linewidth=0.8pt,fillstyle=solid,fillcolor=black](4,6){.15}
\pscircle[linewidth=0.8pt,fillstyle=solid,fillcolor=black](5.75,3.25){.15}
\rput(4,6.5){$\pi$}
\rput(1.25,3.25){$M_1$}
\rput(5.875,4.675){$M$}
\rput(6.3,3.25){$\rho$}
\rput(7.625,1.875){$M_2$}\rput(0.5,0){1}
\rput(7.5,0){2}
\rput(4,-1){\text{rooted at $\pi$}}
\rput(10.25,3){$\cong$}
\psline{->}(16,6)(14.25,3.25)
\psline{-<}(16,6)(16.875,4.675)
\psline{-}(16,6)(17.75,3.25)
\psline{-}(16,6)(12.50,0.50)
\psline{->}(17.75,3.25)(18.625,1.875)
\psline{-}(17.75,3.25)(19.50,0.50)
\pscircle[linewidth=0.8pt,fillstyle=solid,fillcolor=black](16,6){.15}
\pscircle[linewidth=0.8pt,fillstyle=solid,fillcolor=black](17.75,3.25){.15}
\rput(16,6.5){$\pi$}
\rput(13.25,3.25){$M_1$}
\rput(17.875,4.675){$N$}
\rput(18.3,3.25){$\rho$}
\rput(19.625,1.875){$M_2$}\rput(12.5,0){1}
\rput(19.5,0){2}
\rput(16,-1){\text{rooted at $\rho$}}
\endpspicture
\caption{Using the generalized pulley principle.}
\label{pic:twotaxapulley}
\end{figure}

For illustrative purposes we consider the consequence to the stochastic distance of the rerooting of a phylogenetic tree of two taxa. We consider the phylogenetic trees illustrated in Figure \ref{pic:twotaxapulley}, and by using the generalized pulley principle define their respective transition matrices so that their probability distributions are identical: 
\beqn
n_{ij}&=\frac{m_{ji}\pi_{i}}{\rho_j},\nonumber\\
\rho_i&=\sum_j\pi_jm_{ji}.
\eqn
We find that in the first case that we have
\beqn
\omega_\pi(1,2)=-\ln\det M_1-\ln\det M-\ln\det M_2,\nonumber
\eqn
and in the second case
\beqn
\omega_\rho(1,2)=-\ln\det M_1-\ln\det N-\ln\det M_2.\nonumber
\eqn
Now in general $\det M\neq \det N$ and we see that the two possible pairwise distances are not expected to be equal. However, from an empirical perspective it is impossible to distinguish these two possible theoretical scenarios because the probability distributions are identical. Now because any estimator of the pairwise distance must be inferred from the observed distribution, we conclude that one must be careful to consider exactly what theoretical quantity one is obtaining an estimate of. For the case of the $\log\det$ formula we find that quantity it is estimating depends essentially upon the base composition of the observed sequences as follows: 
\\\indent Consider the pairwise distance $d'_{ab}$ given by (\ref{logdetharm}), from the generalized pulley principle we see that this formula will give an estimate of the stochastic distance between $a$ and $b$, where the common ancestral node is placed such that the quantity
\beqn\label{push}
\chi(a,b):=\prod_i\pi^{(a,b)}_i-\left[\prod_{i_1,i_2}\pi^{(a)}_{i_1}\pi^{(b)}_{i_2}\right]^{\frac{1}{2}}
\eqn 
is minimized. Thus the $\log\det$ method will be inconsistent in the sense that, if there has been compositional heterogeneity, the pairwise distance it produces will be an estimate for the edge length assignment where $\chi(a,b)$ is minimized. This may have nothing to do with true placement of the common ancestral vertex and it may even be the case that $\chi(a,b)$ has multiple minimum points. The situation amounts to the fact that, for a given phylogenetic tree, one is (potentially) using the $\log\det$ to estimate pairwise distances with a different edge length assignment for each and every pair of taxa. Clearly for the analysis of multiple taxa this could be become a significant problem and any alternative approach which removes this inconsistency would be beneficial to the analysis. 
\\\indent We see that the consequences of the generalized pulley principle and (\ref{consistency}) to the interpretation of the Markov model of phylogenetics are quite subtle. The generalized pulley principle is telling us that there is no direct way to distinguish the rootedness (and equivalently the directedness of internal edges) of phylogenetic trees. This is due to the fact that there are (infinitely) many phylogenetic trees of identical topology which generate identical probability distributions, differing only by the assignment of stochastic distance and the associated redirection of internal edges.
\indent

\section{The quartet case}

\begin{figure}[t]
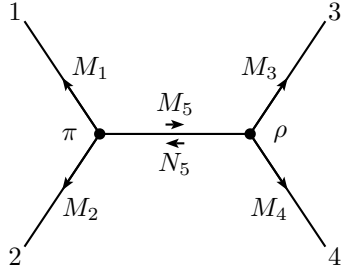

\label{fig:fourtaxa}
\pspicture[](-5.3,)(,)
\psset{linewidth=\pstlw,xunit=0.5,yunit=0.5,runit=0.5}
\psset{arrowsize=2pt 2,arrowinset=0.2}
\psline{-}(2,3)(0,0)
\rput(-.25,-.25){2}
\psline{-}(2,3)(0,6)
\rput(-.25,6.25){1}
\psline{-}(2,3)(6,3)
\psline{-}(6,3)(8,6)
\rput(8.25,6.25){3}
\psline{-}(6,3)(8,0)
\rput(8.25,-0.25){4}
\psline{->}(2,3)(1,4.5)
\rput(1.75,4.75){$M_1$}
\psline{->}(2,3)(1,1.5)
\rput(1.5,1.0){$M_2$}
\psline{->}(6,3)(7,4.5)
\rput(6.25,4.75){$M_3$}
\psline{->}(6,3)(7,1.5)
\rput(6.5,1.0){$M_4$}
\psline{->}(3.75,3.25)(4.25,3.25)
\rput(4,3.8){$M_5$}
\psline{<-}(3.75,2.75)(4.25,2.75)
\rput(4,2.2){$N_5$}
\pscircle[linewidth=0.8pt,fillstyle=solid,fillcolor=black](2,3){.15} 
\pscircle[linewidth=0.8pt,fillstyle=solid,fillcolor=black](6,3 ){.15} 
\rput(1.2,3){$\pi $}
\rput(6.8,3){$\rho$}
\endpspicture 
\caption{Four taxa tree with alternative roots.}
\label{pic:fourtaxapulley}
\end{figure}

\begin{figure}[t]
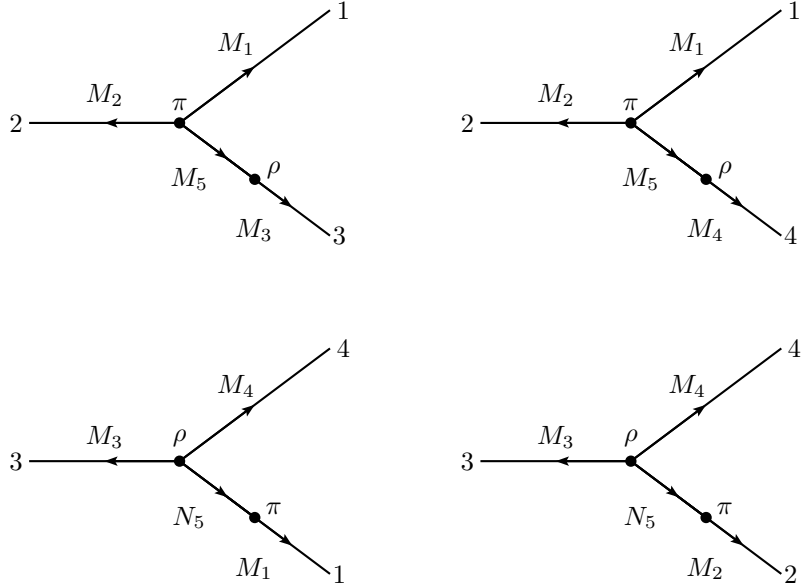

\pspicture[](-2,-5)(,)
\psset{linewidth=\pstlw,xunit=0.5,yunit=0.5,runit=0.5}
\psset{arrowsize=2pt 2,arrowinset=0.2}
\rput(4,3.5){$\pi$}
\psline{-}(4,3)(0,3)
\rput(-.35,3){2}
\psline{-}(4,3)(8,0)
\rput(8.25,0){3}
\psline{-}(4,3)(8,6)
\rput(8.35,6){1}
\psline{->}(4,3)(2,3)
\rput(2,3.75){$M_2$}
\psline{->}(4,3)(6,4.5)
\rput(5.5,5.1){$M_1$}
\psline{-}(4,3)(6,1.5)
\psline{>-}(5,2.25)(6,1.5)
\rput(4.25,1.5){$M_5$}
\rput(6.5,1.75){$\rho$}
\psline{->}(4,3)(7,.75)
\rput(5.969,0.15){$M_3$}
\pscircle[linewidth=0.8pt,fillstyle=solid,fillcolor=black](6,1.5){.15} 
\pscircle[linewidth=0.8pt,fillstyle=solid,fillcolor=black](4,3){.15} 
\rput(16,3.5){$\pi$}
\psline{-}(16,3)(12,3)
\rput(11.65,3){2}
\psline{-}(16,3)(20,0)
\rput(20.25,0){4}
\psline{-}(16,3)(20,6)
\rput(20.35,6){1}
\psline{->}(16,3)(14,3)
\rput(14,3.75){$M_2$}
\psline{->}(16,3)(18,4.5)
\rput(17.5,5.1){$M_1$}
\psline{-}(16,3)(18,1.5)
\psline{>-}(17,2.25)(18,1.5)
\rput(16.25,1.5){$M_5$}
\rput(18.5,1.75){$\rho$}
\psline{->}(16,3)(19,.75)
\rput(17.969,0.15){$M_4$}
\pscircle[linewidth=0.8pt,fillstyle=solid,fillcolor=black](18,1.5){.15} 
\pscircle[linewidth=0.8pt,fillstyle=solid,fillcolor=black](16,3){.15} 
\rput(4,-5.375){$\rho$}
\psline{-}(4,-6)(0,-6)
\rput(-.35,-6){3}
\psline{-}(4,-6)(8,-9)
\rput(8.25,-9){1}
\psline{-}(4,-6)(8,-3)
\rput(8.35,-3){4}
\psline{->}(4,-6)(2,-6)
\rput(2,-5.35){$M_3$}
\psline{->}(4,-6)(6,-4.5)
\rput(5.5,-4){$M_4$}
\psline{-}(4,-6)(6,-7.5)
\psline{>-}(5,-6.75)(6,-7.5)
\rput(4.25,-7.5){$N_5$}
\rput(6.5,-7.25){$\pi$}
\psline{->}(4,-6)(7,-8.25)
\rput(5.969,-8.85){$M_1$}
\pscircle[linewidth=0.8pt,fillstyle=solid,fillcolor=black](6,-7.5){.15} 
\pscircle[linewidth=0.8pt,fillstyle=solid,fillcolor=black](4,-6){.15}  
\rput(16,-5.375){$\rho$}
\psline{-}(16,-6)(12,-6)
\rput(11.65,-6){3}
\psline{-}(16,-6)(20,-9)
\rput(20.25,-9){2}
\psline{-}(16,-6)(20,-3)
\rput(20.35,-3){4}
\psline{->}(16,-6)(14,-6)
\rput(14,-5.35){$M_3$}
\psline{->}(16,-6)(18,-4.5)
\rput(17.5,-4){$M_4$}
\psline{-}(16,-6)(18,-7.5)
\psline{>-}(17,-6.75)(18,-7.5)
\rput(16.25,-7.5){$N_5$}
\rput(18.5,-7.25){$\pi$}
\psline{->}(16,-6)(19,-8.25)
\rput(17.969,-8.85){$M_2$}
\pscircle[linewidth=0.8pt,fillstyle=solid,fillcolor=black](18,-7.5){.15} 
\pscircle[linewidth=0.8pt,fillstyle=solid,fillcolor=black](16,-6){.15}  
\endpspicture
\caption{Three taxon subtrees.}
\label{pic:subtrees} 
\end{figure}

In this section we will show that in the case of a phylogenetic tree of four taxa, the tangle can be used to construct consistent quartet distance matrices. These distance matrices will be consistent in the sense that theoretically they are constructed from one topology with \textit{one} edge length assignment. This should be opposed to the $\log\det$ formula which in the general case can be estimating a different edge length assignment for each and every pairwise distance. 
\\\indent
For analytic purposes we use the generalized pulley principle to root the four taxon tree in two ways, as illustrated in Figure \ref{pic:fourtaxapulley}. The difference between the two cases is simply in the directedness of the internal edge and the generalized pulley principle allows us to calculate the required transition probabilities so that the two trees generate identical probability distributions. The pattern probabilities for the two cases are given by
\beqn\label{fourtaxapulley}
p_{i_1i_2i_3i_4}&=\sum_{j,k}\pi_jm^{(5)}_{jk}m^{(1)}_{ji_1}m^{(2)}_{ji_2}m^{(3)}_{ki_3}m^{(4)}_{ki_4}\\
&=\sum_{j,k}\rho_jn^{(5)}_{jk}m^{(1)}_{ki_1}m^{(2)}_{ki_2}m^{(3)}_{ji_3}m^{(4)}_{ji_4}
\eqn
where to ensure the equality of the two expressions we have
\beqn
n^{(5)}_{ij}=\frac{m^{(5)}_{ji}\pi_{i}}{\rho_j},\nonumber
\eqn
and $\rho_i=\sum_j\pi_jm^{(5)}_{ji}$.
\\\indent
From these expressions we wish to calculate the theoretical values of the formula (\ref{dist}) for each possible group of three taxa. To obtain these values one simply chooses the form of the tree such that after the deletion of a fourth taxon one is left with a three taxon tree of star topology. By sequentially deleting one taxon at a time we are led to the four star topology subtrees illustrated in Figure \ref{pic:subtrees} and the corresponding pattern probabilities are given by the expressions
\beqn
p^{(123)}_{ijk}&=\sum_{l_1,l_2}\pi_{l_1}m^{(1)}_{l_1i}m^{(2)}_{l_1j}m^{(5)}_{l_1l_2}m^{(3)}_{l_2k},\\\nonumber
p^{(124)}_{ijk}&=\sum_{l_1,l_2}\pi_{l_1}m^{(1)}_{l_1i}m^{(2)}_{lj}m^{(5)}_{l_1l_2}m^{(4)}_{l_2k},\\
p^{(134)}_{ijk}&=\sum_{l_1,l_2}\rho_{l_1}n^{(5)}_{l_1l_2}m^{(1)}_{l_2i}m^{(2)}_{l_1j}m^{(4)}_{l_1k},\\
p^{(234)}_{ijk}&=\sum_{l_1,l_2}\rho_{l_1}n^{(5)}_{l_1l_2}m^{(2)}_{l_2i}m^{(3)}_{l_1j}m^{(4)}_{l_1k}.\\
\eqn
From this it is easy to calculate the values simply by considering the results of the previous section:
\beqn\label{fourdelta}
\Delta_{12}^{(3)}&=\omega(1,2),\qquad 
&\Delta_{12}^{(4)}&=\omega(1,2),\\
\Delta_{13}^{(2)}&=\omega_\pi(1,3),\qquad 
&\Delta_{13}^{(4)}&=\omega_\rho(1,3),\\
\Delta_{14}^{(2)}&=\omega_\pi(1,4),\qquad 
&\Delta_{14}^{(3)}&=\omega_\rho(1,4),\\
\Delta_{23}^{(1)}&=\omega_\pi(2,3),\qquad 
&\Delta_{23}^{(4)}&=\omega_\rho(2,3),\\
\Delta_{24}^{(1)}&=\omega_\pi(2,4),\qquad 
&\Delta_{24}^{(3)}&=\omega_\rho(2,4),\\
\Delta_{34}^{(1)}&=\omega(3,4),\qquad 
&\Delta_{34}^{(2)}&=\omega(3,4).
\eqn
where
\beqn
\omega(a,b)&=\omega_a+\omega_b,\\
\omega_\pi(a,b)&=\omega_a+\omega_m+\omega_b,\\
\omega_\rho(a,b)&=\omega_a+\omega_n+\omega_b,\\
\omega_m&=-\ln\det M,\\
\omega_n&=-\ln\det N;
\nonumber
\eqn
and we have made use of (\ref{consistency}) in the form
\beqn
\omega_n=\omega_m-\sum_i(\ln\pi_i-\ln\rho_i).\nonumber
\eqn
\indent
We see that for any two taxa we have two options for assigning a pairwise distance. In the cases of the pairs $(12)$ and $(34)$ we see that either choice is consistent with the other, whereas in the case of the pair $(13)$, $(14)$, $(24)$ and $(34)$ the two choices lead to an inconsistent assignment of the internal edge length upon the tree. Effectively what is happening here is that for a four taxa tree there are two possible edge length assignments for the internal edge and for a given pair of taxa $(ab)$ and third taxa $c$, the tangle formula (\ref{dist}) is estimating the distance between $a$ and $b$ by assigning one of the two possible edge lengths to the internal edge depending on the topology of the tree.
\\\indent 
It is possible to eliminate this inconsistency by using either a \textit{max} or \textit{min} criterion in the construction of the distance matrix:
\beqn
\phi^{max}_{ab}:=max\{\Delta_{ab}^{(c)},\Delta_{ab}^{(c')}\}\nonumber
\eqn
or
\beqn
\phi^{min}_{ab}:=min\{\Delta_{ab}^{(c)},\Delta_{ab}^{(c')}\}.\nonumber
\eqn
By making one of these choices to construct a distance matrix one has choosen the directedness of the internal edge of the phylogenetic tree (\ref{fig:fourtaxa}) consistently. This procedure leads to an improvement of consistency upon the $\log\det$ technique for the construction of quartet phylogenetic distance matrices. It is hoped that this technique can be used fruitfully to improve the reconstruction of phylogenetic quartets, which can be used as a first step in the reconstruction of large phylogenetic trees \cite{bryant2001,strimmer1996}.
\\\indent

\section{Conclusion}
In this paper we have given a review of the standard assignment of branch weights to phylogenetic trees, reviewed the use of the $\log\det$ formula as an estimator of pairwise distances and shown how a previously unknown polynomial, the tangle, can be used to construct an improved estimator. We have generalized Felsenstein's \textit{pulley principle} and used this result to show exactly how the distance matrix estimates become inconsistent when applied to the reconstruction problem of multiple taxa. We have shown that the tangle formula along with a \textit{max/min} criterion can be used to remove this inconsistency and construct consistent quartet distance matrices. 

\section{Aknowledgements}
The authors would like to thank Jim Bashford and Patrick McLean for assistance with the computational aspects of this work; Alexei Drummond, Malgorzata O'Reilly, Michael Charleston and Lars Jermiin for helpful comments; and the Antarctic Climate and Ecosystems Cooperative Research Centre for use of their computational facilities. This research was supported by the Australian Research Council grant DP0344996 and the Australian Postgraduate Award.

\end{document}